# Mr. Doc: A Doctor Appointment Application System

Shafaq Malik, Nargis Bibi, Sehrish Khan, Razia Sultana, Sadaf Abdul Rauf
Department of Computer Science
Fatima Jinnah Women University
Rawalpindi, Pakistan
Shafaqmalik94@gmail.com, nargis@fjwu.edu.pk, sehrishkhan693@gmail.com, razia141@hotmail.com, sadaf.abdulrauf@gmail.com

*Abstract*--Life is becoming too busy to get medical appointments in person and to maintain a proper health care. The main idea of this work is to provide ease and comfort to patients while taking appointment from doctors and it also resolves the problems that the patients has to face while making an appointment. The android application Mr.Doc acts as a client whereas the database containing the doctor's details, patient's details and appointment details is maintained by a website that acts as a server.

## I. INTRODUCTION

If anybody is ill and wants to visit a doctor for checkup, he or she needs to visit the hospital and waits until the doctor is available. The patient also waits in a queue while getting appointment. If the doctor cancels the appointment for some emergency reasons then the patient is not able to know about the cancelation of the appointment unless or until he or she visits the hospital. As the mobile communication technology is developing rapidly, therefore, one can use the mobile's applications to overcome such problems and inconvenience for the patients. There is much work in the literature in this regard [1-14]. An intelligent agent based appointment system has been proposed in [1] in which a scheduling system is provided for patients. The junior medical staff schedules appointment according to the priority level. [2] proposed an Android application that is used to remind the patients of their dosage timings through Alarm Ringing system so that they can stay fit and healthy. Searching doctors and hospitals alongwith navigation details are also available in the app so they can get proper treatment on time. [3] proposed an android based appointment management system which uses application programming interfaces (APIs) from Google map and calendar. This appointment based application can be used with other appointment based systems. The mobile application accepts appointments by saving the record of the appointment on the phone calendar which is synchronized with the Google calendar. The user gets an alert based on preset specified time before the appointment time and date. [4] proposed a Health Track system that communicates with sensors via smart phone for data collection, and stores data concurrently to the central server for further analysis via the internet. Some online systems that are already functional still have some drawbacks. To overcome these drawbacks an online patient appointment system is proposed using Near Field Communication (NFC) technique and Android enabled mobile application. This system works by registration and scheduling appointments based on NFC that accesses patient's health records and reports to alert nurses and doctors. There is another interesting work which is Disease Self-inspection and Hospital Registration Recommendation System (DSRRS) [6]. It uses Representational State Transfer (REST) style for communication interface between reasoning service and the system. Before reasoning users disease history is retrieved from Personal Health Record (PHR) and passed as an input to reasoning service. Mainly the input contains User's information, disease history, Knowledge base (symptoms) and output of reasoning service. [7] described an android smart phones and tablets application that is freely downloadable from Google play store and it provides various functionalities including personnel medical records, to trace position of actual user in real-time. Routing algorithm is used to find minimum distance for destination building. Another study consists of an online database for the monitoring of patient with artificial heart [8]. This database consists of monitoring terminal that is portable and keeps continuous record of a patient including history. There are other studies which involve handheld healthcare [9, 10, 11] and efficient algorithms for appointment scheduling including self-inspection [12, 13, 14].

The proposed work in this paper is an Online Hospital Management Application that uses an android platform that makes the task of making an appointment from the doctor easy and reliable for the users. Android based online doctor appointment application "Mr. Doc" contains two modules. One module is the application designed for the patient that contains a login screen. The patient has to register himself before logging in to the application. After logging in, the patient can select a hospital and can view the hospital details. The patient has the option of selecting a doctor from the list of doctors and can view the doctor's details. The patient can request for an appointment on his/her preferred day/time. The selected day/time slot will be reserved and patient will receive the notification of the successfully added appointment. The patient can view the location of the hospital on map. In addition, the patient can contact to the hospital and the doctor by making a call or may send an email to the doctor.

The second module is the admin module that is designed on the website. The admin views all details of doctors and all appointments by the admin. The admin can add doctor, view patient's details and doctor's details and can view appointments also. All the doctors of the specific clinic are registered by the admin. Doctors cannot register themselves.

Rest of the paper is organized as follows. Section II explains the design interface and the tools which have been used. Section III discusses the implementation and screenshots. Section IV concludes the paper.

## II. DESIGN INTERFACE

The front end design is simple and user-friendly. Once the application is started the patient will register himself and then he will be able to log in into the application. The patient can make an appointment by selecting the preferred doctor, date and time. The appointments are managed by the admin through a website. The admin also registers a doctor. Admin is able to view doctors, view patient's records and view feedback also. The back end design includes a server which acts as a centralized database. All the data of the registered doctors and patients and the data regarding the appointments are placed on the server. The data is approached and shared by using API'S between the website and the android application.

### A. ANDROID

Android is an open source operating system which is Linux based and android platform is used to develop many useful applications for the mobile devices that makes the tasks of everyday life easy and faster. The android platform also provides built in database (SQLite database) and Web services. Android platform provides connectivity between the server and the application using certain APIs, hence the task of making a doctor appointment using a mobile application connected to a website located on the server becomes easy using the advanced features and libraries available on the android platform.

### B. SOFTWARE DEVELOPMENT TOOLS

The following software tools were used during the development process.
- Android studio 2.1.1 and SDK plug-in
- JDK 6
- Android 6.0 (Marshmallow) installed packages
- Ipage Server
- HTML
- Php

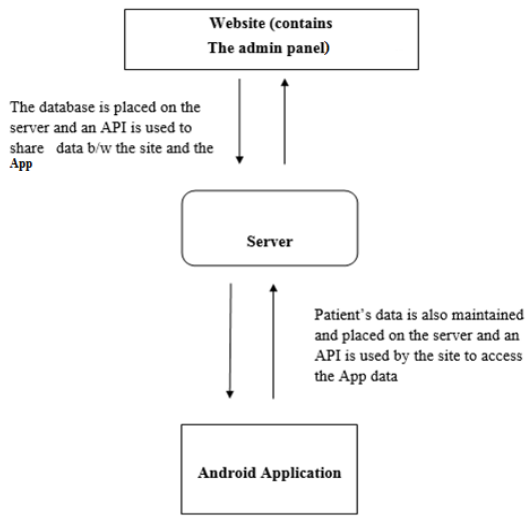

Figure 1: Block diagram of the appointment application

### III. IMPLEMENTATION

The user will firstly downloads the application and install it in their mobile devices. Figure 1 shows the block diagram of the application. Once installed, this application will remain into the device permanently until the user deletes it or uninstalls it. After the installation when the user clicks on the app icon, the first thing that will appear on the screen is splash screen that contains the application's logo as shown below

Figure 1 *showing the splash screen*

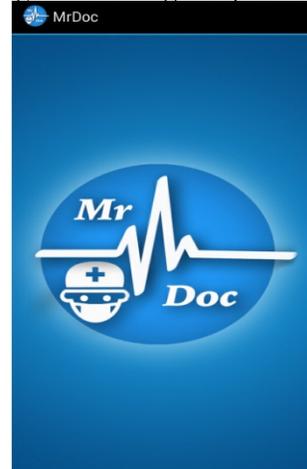

The patient will have to register in the application on first use. After registration, the patient will receive a username and password. For sign up, the user has to fill the given fields that are username, email, password and confirm password and then the user clicks on the signup button to register itself and then all the information provided by the user is saved in the database located on the server. The signup screen is shown below

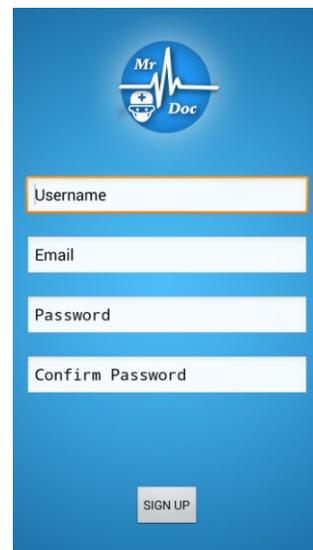

Figure 2: The Signup screen

If the user registers successfully then a notification message "successfully registered" is displayed as shown below

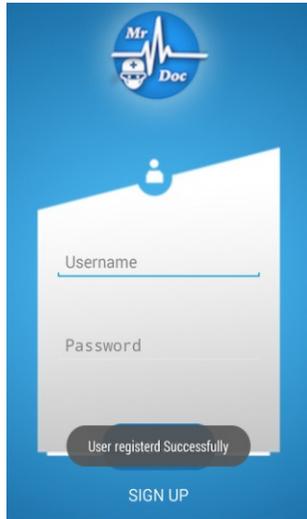

Figure 3: User registered successfully

Different checks are also maintained while registering the user. If both the passwords are not matched then the user will be notify that the "passwords didn't matched" as shown below

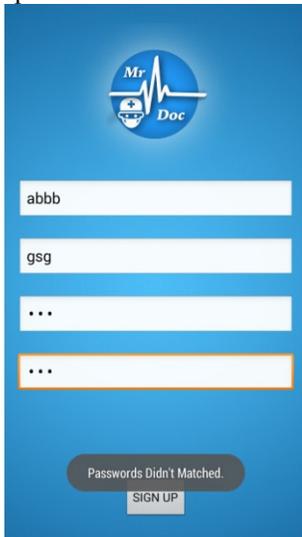

Figure 4: Passwords didn't match

And if email is not valid then the user cannot register itself and a notification will displayed that "email is not valid" as shown below

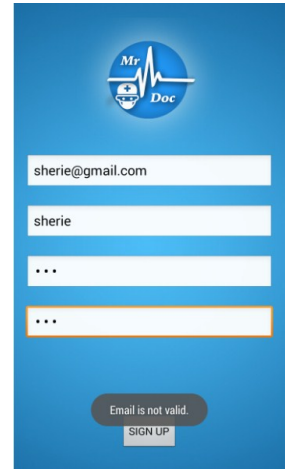

Figure 5: Email is not valid

The patient has to use this username and password for logging into the app for each time usage. For signing in the user has to provide the registered username and password otherwise if the user enters such a username or password that is not registered then the user will get a notification message that "Signin failed check your connection or contact support" as shown below

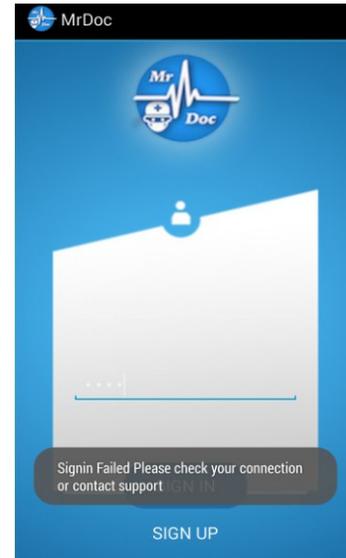

Figure 6: Sign in failed

After logging in, the menu screen is displayed containing different option like hospitals, doctors, health schedules, about and sign out as shown below

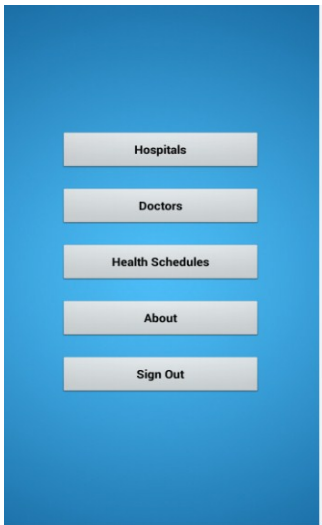

Figure 7: The menu screen

If the patient selects the hospitals option then he/she can view a list of hospitals. Then the patient selects the particular hospital and then he can view the hospital details as shown below

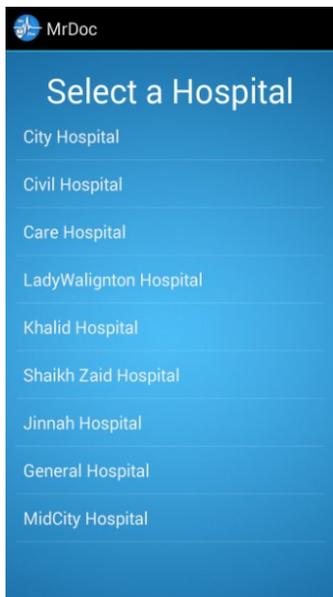

Figure 8: The list of hospitals

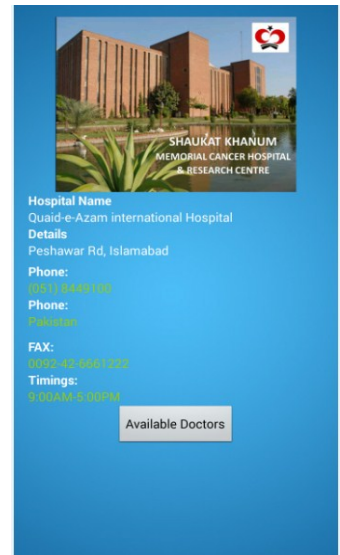

Figure 9: The hospital details

Patient can contact to the hospital by making a call by clicking on the hospital phone number as shown below

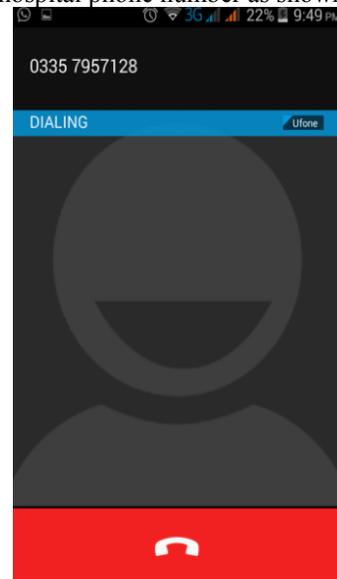

Figure 10: The user calling to hospital

The patient can select any particular doctor and view his profile by clicking on the available doctors or by selecting the doctor's option from the menu screen. The list of the available doctors is displayed as shown below

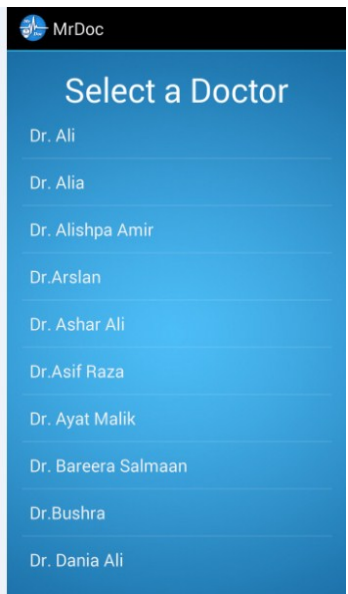

Figure 11: The list of doctors in the selected hospital

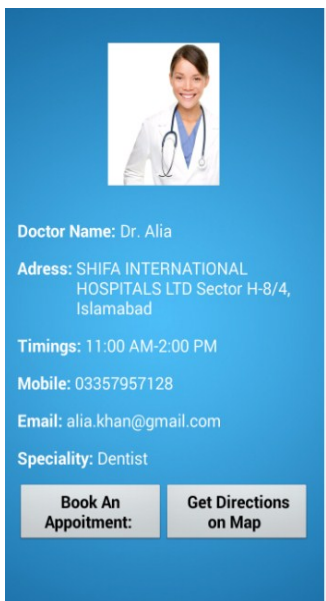

Figure 12: The selected doctor details

Patient can contact the doctor by making a call by clicking on the doctor's phone number or patient can also sends an email by clicking on the doctor's email as shown below

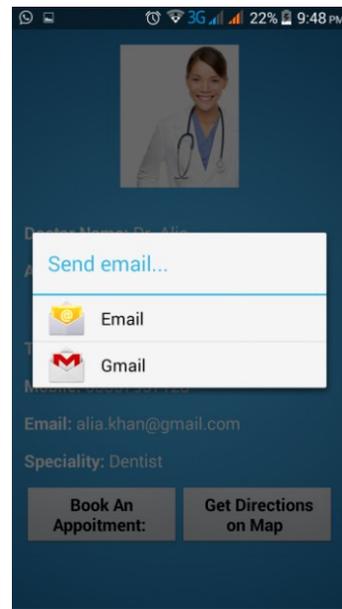

Figure 13: The user sending email to doctor

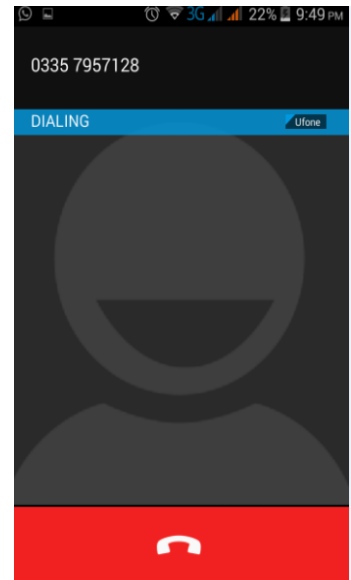

Figure 14: The user calling the doctor

By clicking in the book appointment button, a calendar and different available time slots are displayed on the screen. The patient has to send a request for appointment by selecting a day or time. The central database gets updated accordingly. The user will get notification message of "successfully added" if the appointment is successfully registered in the database as shown below

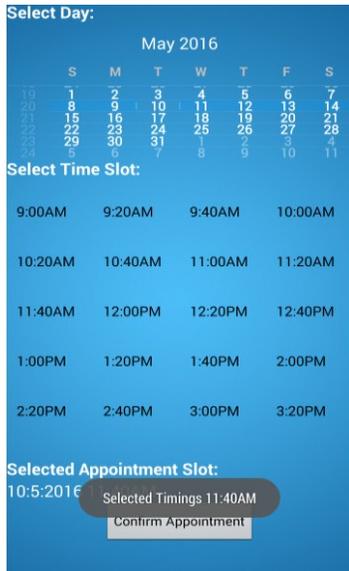
Figure 15: The selected day & time slot

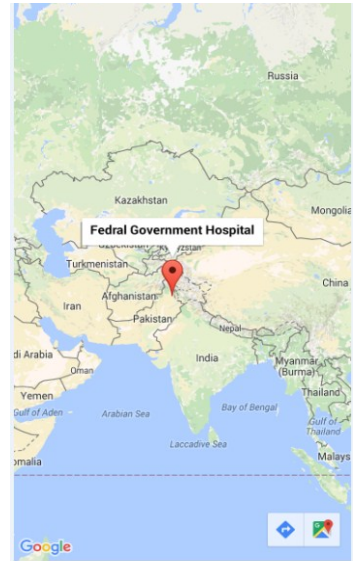
Figure 17: The user can see the hospital location

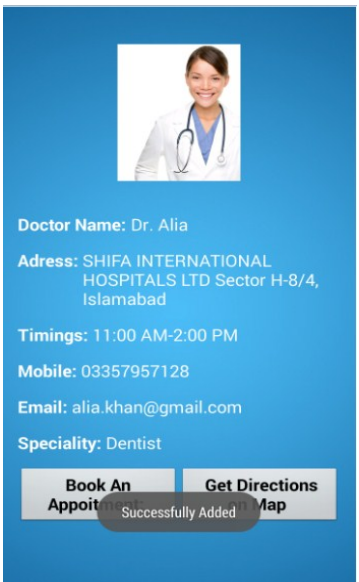
Figure 16: The appointment reserved successfully

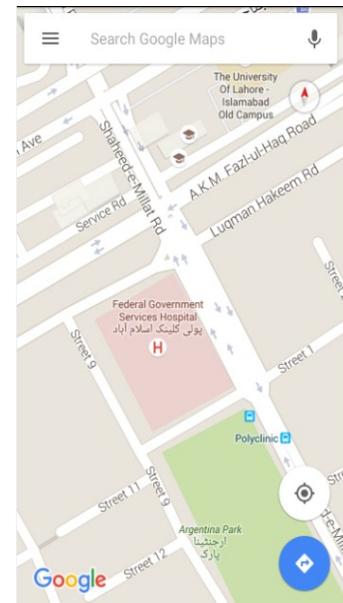
Figure 18: Showing the hospital location

By clicking on the "get directions on map" button, the location of the hospital is displayed on the screen using the Google maps as shown below

By clicking on the health schedule option, the screen containing different health schedules of different age groups is displayed as shown below

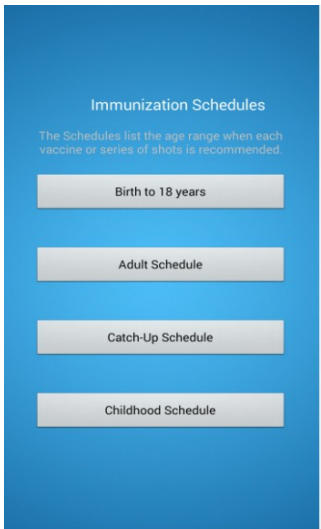

Figure 19: The different options of health schedules

By selecting the particular option, the health schedule of that age group is displayed i.e if user selects the childhood schedule then it is displayed as shown below

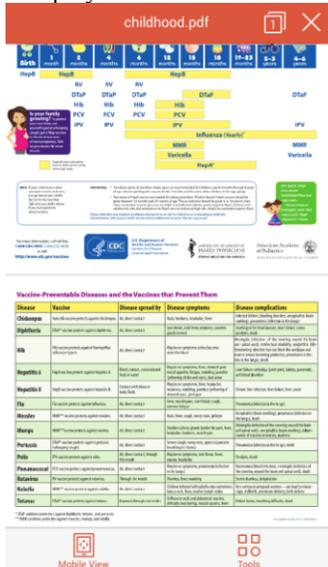

Figure 20: Showing the health schedule

The About option on the menu screen shows the application's objectives and the developers of the application is displayed as shown below.

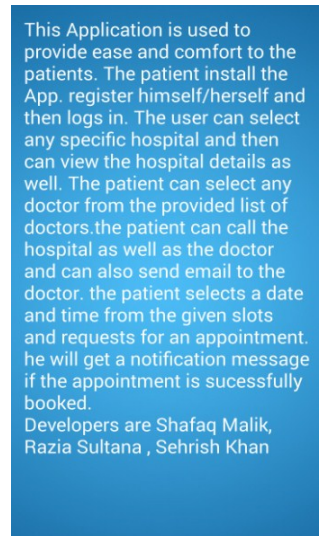

Figure 21: The About screen

The patient gets logged off by clicking on the sign out button on the menu screen. Another module of this system is located on the website. Admin logs in and then he can add a doctor. Admin can view the doctor details, patient details and the appointment details. Data is shared between both the website and application by using APIs. The screen shots of the website are shown below

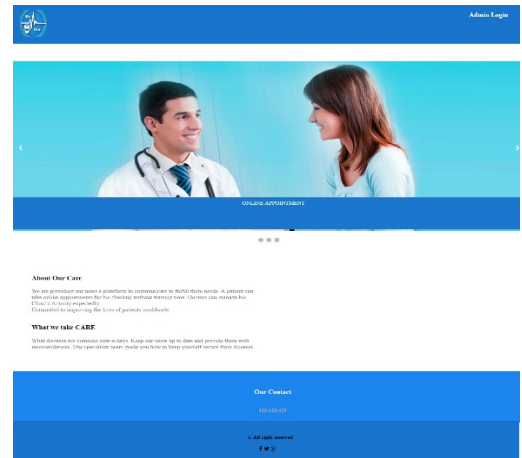

Figure 22: The main page of the website

Figure 23: The login page for admin

Figure 24: The admin portal

Figure 25: The form to add a new doctor

Figure 26: Showing the details of added doctors

Figure 27: Showing the details of registered patients

Figure 28: Showing the appointments of the patients

## IV. CONCLUSION AND FUTURE WORK

The proposed online appointment system has been implemented in android studio for application development and website is developed using HTML and PHP. The tasks involved in this work are divided into modules. The data is approached and shared by using API'S between the website and the android application. The proposed system is efficient and has friendly user interface. Addition of the admin and doctor modules in the android application are included in future work. That would help the doctor to register on the application and perform all the tasks on the app. The admin would be able to use the app for managing the details of the patients and the doctors instead of using the website. A payment or some amount may be charged to the users/patients while making an appointment to avoid the unethical users. As many users only register themselves just for fun and has no concern by making an appointment. Some more future directions are the improvements in the patient's module which includes setting reminders for the appointments and saving the appointment date to the calendar.